\documentclass[10pt]{revtex4}
\usepackage{hyperref}
\usepackage{amsmath}
\usepackage[psamsfonts]{amssymb }
\usepackage{mathrsfs}
\allowdisplaybreaks
\begin{document}
\title{\Large Hamiltonian and Lagrangian BRST Quantization in Riemann Manifold II}

\author{Vipul Kumar Pandey\footnote {e-mail address: vipulvaranasi@gmail.com}}

\affiliation {Department of Physics, Chandigarh University, Mohali
- 140413, INDIA.}

\begin{abstract}
We have previously developed the BRST quantization on the hypersurface $V_{N-1}$ embedded in N dimensional Euclidean space $R_N$ in both Hamiltonian and Lagrangian formulation. We generalize the formalism in the case of L dimensional manifold $V_L$ embedded in $R_N$ with $1\leq L < N$. The result is essentially the same as the previous one. We have also verified the results obtained here using a simple example of particle motion on a torus knot.
\end{abstract}

\maketitle
\section{Introduction}\label{Introduction}
In the previous work \cite{VKP} we have considered the BRST quantization of the motion on a hypersurface $V_N$ embedded in the N dimensional Euclidean space $R^N$ based on Batalin-Fradkin-Fradkina-Tyutin (BFFT) Abelianized Batalin-Fradkin-Vilkovisky (BFV) and Batalin-Vilkovisky (BV) formalisms. It is a well known result in differential geometry that any $N$ dimensional Riemann manifold can be locally embedded in the $\frac{N(N+1)}{2}$ dimensional Eucliden space but can not be embedded in $(N+1)$ dimension generally \cite{YAS}. In this sense, the considerations in the previous manuscript \cite{VKP} should be extended to general Riemannian manifolds. This is the motivation of the present manuscript. 

It is well known in the literature that the quantization of the system in curved space has been extensively studied about the ordering problem using two different approaches, canonical and path integral \cite{HEWCRS,RS,TK,TKRS,TORS1,TORS2,TKTORS,MOHS,FSTK,FKST,TKW,HKTK,JLGAN,MO,BSKK}. At the same time, the quantization of dynamical systems constrained to curved manifolds embedded in the higher-dimensional Euclidean space has been extensively investigated as one of the quantum theories \cite{DWOS,AFFR,NDBPD,LPDT,IBMGSL1,IBMGSL2,IGSLAS}. Here, a non-relativistic particle constrained to move on a curved surface embedded in the higher dimensional Euclidean space \cite{NOKFAK,NOKFCK} has been taken. These systems and their various properties they possess, have been investigate by many authors \cite{NOG,INTT,STTT,CFFM,NO,FOUC,CFK,AS,LHL,NOG1,NOG2,NOMN,DPO,FGK,AVG,KFNO,NOM1,NOM2,MN,PM1,PM2,NCAR1,NCAR2,NCAR3,DOBSRT}.  This has motivated us to extend our previous work\cite{VKP} to more general class of systems discussed in \cite{NOKFCK}. 

                                    Becchi-Rout-Stora-Tyutin (BRST) quantization
\cite{CRS1,CRS2,CRS3,IVT} is on of the most significant technique to deal with a system with constraints. It has also been found to be symmetry of general class of constrained systems \cite{PAMD1,JLAPGB,PAMD2,KS,MHCT,MH}. In this quantization method, we enlarge the total Hilbert space of the gauge system under study and bring back the gauge symmetry of the gauge fixed action in the extended phase space, keeping the physical contents of the theory unchanged. BRST symmetry plays very significant role in renormalization of spontaneously broken gauge theories like, standard model and hence is of very high significance for different kind of systems. To the best of our knowledge, there is no literature available which studied the BRST symmetry for a particle moving on a hypersurface $V_L (1\leq L < N)$ embedded in the Euclidean space $R_N$. This motivates us in the study of BRST symmetry for this system. We will do the constraints analysis of this sytem using the Dirac's technique. The system is shown to contain second-class constraints. The BFFT method will be used to convert the second class constraints to first class one \cite{LFSS,IABESF1,IABESF2,BFFR,IABIVT,EERM,RBBN,SG,CBAPMT,IBVKAR,ILBAAR,AAR,IBAR,VPRT,MPT}. Then the BRST charge and symmetries will be constructed for this BFFT Abelianized system using BFV method \cite{ESFGV,IABGV,IABESF} with the help of Faddeev-Senjanovic technique \cite{LDFAD,PSENJ}. In the limit $L\rightarrow (N-1)$ the system will return to sytem in \cite{VKP}. The results developed in the manuscript have been verified using a model of particle on torus knot \cite{VS,PDSG,VKPBPM,VPBM,ASSG}. At the end BV-BRST quantization of the BFFT Abelianized sytem will be investigated \cite{IABGAV1,IABGAV2,JGJPSS,RART}. Recently Lagrangian Abelianization procedure for constrained systems has also been developed \cite{SLL, VAASLL}.

This is the second and final part of the work. In this part we will discuss BRST quantization of embedding $V_L$ in Euclidean space $R_N$ where $1\leq L < N$. The paper has been organized in the following way. In the second section, we have reviewed motion in curved space and also calculated all the possible constraints of the theory using Dirac's constraints analysis method. In the third section, we have reviewed BFFT formalism. In the section four, we have constructed first class constraints and Hamiltonians for the general class of systems under study. In the next section we will construct BRST symmetry for this system based on BFV Formalism. In the section six, we have shown consistency of our results in the limit $L \rightarrow (N-1) $. In the section seven, we have given a simple example of this kind of systems. In the section eight, we have discussed BV quantization of this system based on BFFT formalism. In the next section concluding remarks have been made. In the end we have discussed some important calculations in the appendix.

\section{Classical Mechanics on $V_L$ in $R^N$}
\label{Classical Mechanics on $V_L$ in $R^N$}
Let us consider an N dimensional Euclidean space ${R_N}$, a point in which is specified by a set of Cartesian coordinates 
\begin{eqnarray}
X^A : \{x^1,x^2,...,x^N\}.
\end{eqnarray}

Further consider in $R_N$ an $L$ dimensional Riemann subspace, $V_L (1\leq L < N)$, a point of which is specified by a set of coordinates $q^k$ \cite{NOKFCK},
\begin{eqnarray}
q^k :  \{ q^1, q^2, ..., q^L\}
\label{qkgnc}
\end{eqnarray}
The metric of this system is defined as $g_{ij} (q^k)$. We can construct in $R_N$ a set of curvilinear coordinates including $(q^1, q^2, ..., q^L)$,
\begin{eqnarray}
q^{\mu} :  \{ q^1, q^2, ..., q^L, q^{L+1}, ..., q^N\}
\label{qalp}
\end{eqnarray}
Let us assume that $\{q^{L+a}\} (a: 1\sim {N-L})$ are the intrinsic coordinates normal to $V_L$ \cite{NOKFCK}. We can also use the notation
\begin{eqnarray}
Q^a \equiv q^{L+a}, \quad  a: 1 \sim {N-L}
\label{qal}
\end{eqnarray} 
Then the subspace $V_L$ can be defined as
\begin{eqnarray}
Q^a = q^{L+a} = 0
\label{Qa}
\end{eqnarray}
The metric for the curvilinear coordinates $q^{\mu}$ in $R_N$ is defined as
\begin{equation}
{\tilde g}_{\mu\nu} = 
\begin{pmatrix}
{\tilde g}_{ij} & N_{ib} \\
N_{ja} & {\tilde G}_{ab}\\
\end{pmatrix}
\label{tigmn}
\end{equation}
where $N_{ia}$ and ${\tilde G}_{ab}$ are defined as 
\begin{eqnarray}
N_{ia} \equiv {\tilde g}_{\mu = i, \nu = L+a}, \quad {\tilde G}_{ab} \equiv {\tilde g}_{\mu = L+a, \nu = L+b}
\label{NLGg}
\end{eqnarray} 
It is worth notice that the metric $g_{ij} = {\tilde g}_{ij} (q^k, Q^a = 0)$ is induced metric on $V_L$ and the metric $G_{ab} = {\tilde G}_{ab} (q^k, Q^a = 0)$ can be defined as some function on $V_L$. Using this assumption, we will get $N = 0$ when $Q = 0$. So the metric on $V_L$ have form
\begin{equation}
g_{\mu\nu} = 
\begin{pmatrix}
{\tilde g}_{ij} & 0 \\
0 & {\tilde G}_{ab}\\
\end{pmatrix}
\label{gGmn}
\end{equation}    
and the inverse matrix is defined as
\begin{equation}
g^{\mu\nu} = 
\begin{pmatrix}
g^{ij} & 0 \\
0 & G^{ab}\\
\end{pmatrix}
\label{invgmn}
\end{equation}
which implies that
\begin{eqnarray}
g^{\mu\zeta}.g_{\zeta\nu} = {\delta^\mu}_\nu
\label{gmznd}
\end{eqnarray} 
which can further be written as, 
\begin{eqnarray}
g^{ij}\cdot g_{jk} = \delta^i_k, \quad G^{ab}\cdot G_{bc} = {\delta^a}_c
\label{gijg}
\end{eqnarray} 
We know that ${\tilde g}_{\mu\nu}$ can also be written as
\begin{eqnarray}
{\tilde g}_{\mu\nu} = (\frac{\partial x}{\partial q^\mu})\cdot (\frac{\partial x}{\partial q^\nu}).
\label{qmqn}
\end{eqnarray} 
From here, we can obtain following relations \cite{NOKFCK}
\begin{eqnarray}
0 &=& (\frac{\partial x}{\partial q^k})\cdot (\frac{\partial x}{\partial Q^a})|_{Q = 0} = \sum_{A} e_k^A\cdot h_a^A\nonumber\\
g_{ij}(q^k) &=& (\frac{\partial x}{\partial q^i})\cdot (\frac{\partial x}{\partial q^j})|_{Q = 0} = \sum_{A} e_i^A\cdot e_j^A\nonumber\\
G_{ab}(q^k) &=& (\frac{\partial x}{\partial Q^a})\cdot (\frac{\partial x}{\partial Q^b})|_{Q = 0} = \sum_{A} h_a^A\cdot h_b^A
\label{gGQqA}
\end{eqnarray} 
where $e_i^A$ and $h_a^A$ are defined as
\begin{eqnarray}
e_i^A (q^k) \equiv (\frac{\partial x^A}{\partial q^i})|_{Q = 0}, \quad h_a^A (q^k) \equiv (\frac{\partial x^A}{\partial Q^a})|_{Q = 0}
\label{eAqk}
\end{eqnarray}
Here, $e_i^A$ are called the natural frame and gives the induced metric on $V_L$. 
The inverse metric of ${\tilde g}_{\mu\nu}$ is given by  
\begin{eqnarray}
{\tilde g}^{\mu\nu} = \nabla q^{\mu}\cdot \nabla q^{\nu}
\label{nqmnqn}
\end{eqnarray} 
where $\nabla \equiv \frac{\partial}{\partial x}$.
From this, we obtain
\begin{eqnarray}
g^{\mu\nu} = {\tilde g}^{\mu\nu}|_{Q = 0} = \nabla q^{\mu}\cdot \nabla q^{\nu}|_{Q = 0}.
\label{tignq}
\end{eqnarray} 
 
The Lagrangian for the particle motion on $V_L$ is defined as \cite{NOKFCK},
\begin{eqnarray}
L = \frac{1}{2}\cdot{\dot x}^A{\dot x}_A - V(x) + \lambda_a Q^a(x)
\label{Lagpm}
\end{eqnarray}
Here $``A"$ varies between 1 to $N$ and $``a"$ from 1 to $N-L$. The metric for the coordinate $x^A$ is $\delta_{AB}$. $\lambda_a'S$ are variables independent of $x^A$ and the dot denotes the time derivative.
The canonical momentum conjugate to $x^A$ and $\lambda^a$ can be written as
\begin{eqnarray}
P_A &\equiv& \frac{\partial L}{\partial {\dot x}^A} = {\dot x}^A \nonumber\\
\Pi^a &\equiv & \frac{\partial L}{\partial {\dot \lambda}_a} \approx 0
\label{PAPia}
\end{eqnarray}
Hamiltonian corresponding to Lagrangian in eqn (\ref{Lagpm}) can be written as,
\begin{eqnarray}
H _0 = \frac {1}{2}\cdot P_A P^A + V(x) - \lambda_a Q^a(x)
\label{H0}
\end{eqnarray}

\subsection{Hamiltonian Analysis}\label{Hamiltonian Analysis} 
The primary constraint for the system under study is defined as,
\begin{equation}
\Pi_a \approx 0
\label{pia}
\end{equation} 
After including the primary constraint, the new Hamiltonian is written as, 
\begin{equation}
H_T = \frac {1}{2}\cdot P_A P^A + V(x) - \lambda_a Q^a(x) + u_a \Pi^a
\label{HT}
\end{equation}
where $u_a'S$ are a set of Lagrange multipliers for the system.
Now, we will perform the constraint analysis of the given system using the Dirac's technique of constraints analysis \cite{PAMD1,JLAPGB,PAMD2,KS,MHCT,MH}. All the constraints of the theory can be calculated in the following manner \cite{NOKFCK},
\begin{eqnarray}
{\dot \Pi}^a & = & \{ \Pi^a, H_T \}_P = Q^a \nonumber\\
{\ddot \Pi}^a & = & \{ Q^a, H_T \}_P = P^A \cdot {\partial_A Q^a} \nonumber\\
{\Pi^a}^{(3)} &=& \{ DQ^a, H_T \}_P = D^2Q^a - \nabla Q^a\cdot \nabla (V - \lambda_d Q^d) 
\label{ddPia}
\end{eqnarray}
The constraint $\Pi^{a(4)} = 0$ determines the $u_a'S$ and the procedure is over. So, the explicit form of the constraints are,
\begin{eqnarray}
\Phi_1^a &=& \Pi^a \approx 0 \nonumber\\
\Phi_2^a &=& Q^a \approx 0 \nonumber\\
\Phi_3^a &=& DQ^a \approx 0 \nonumber\\
\Phi_4^a &=& P^AP^B\partial_A\partial_B Q^a  - \nabla Q^a\cdot{\nabla}(V - \lambda_d Q^d) = D^2 Q^a - \nabla Q^a \cdot{\nabla}\Phi \approx 0 
\label{Phia}
\end{eqnarray}
Here D and $\Phi$ are defined as, $D = P^A \partial_A$, $\Phi = (V - \lambda_a Q^a)$. Also, the product of partial derivatives is defined as, $\nabla f\cdot\nabla g \equiv \sum_A \partial_A f\cdot \partial_A g$. 

The Poisson brackets between the constraints are defined as \cite{NOKFCK},
\begin{eqnarray}
\{\Phi_1^a, \Phi_4^b \}_P &=& - \nabla Q^a\cdot \nabla Q^b \equiv - \alpha^{ab}, \nonumber\\
\{\Phi_2^a, \Phi_3^b \}_P &=& \nabla Q^a\cdot \nabla Q^b \equiv \alpha^{ab},\nonumber\\
\{\Phi_2^a, \Phi_4^b \}_P &=& 2\nabla Q^a\cdot (\nabla DQ^b)\equiv -\beta^{ab}, \nonumber\\
\{\Phi_3^a, \Phi_4^b \}_P &=& 2\nabla (D Q^a)\cdot \nabla (D Q^b) - \nabla Q^a\cdot \nabla \Phi_4^b \equiv - \gamma^{ab} ,\nonumber\\
\{\Phi_3^a, \Phi_3^b \}_P &=& \nabla (DQ^a)\cdot \nabla Q^b - \nabla Q^a\cdot \nabla (DQ^b)\equiv \rho^{ab}\nonumber\\
\{\Phi_4^a, \Phi_4^b \}_P &=& 2\left[\nabla \Phi_4^a\cdot \nabla (DQ^a) - \nabla \Phi_4^b\cdot \nabla (DQ^a)\right]\equiv \epsilon^{ab}
\label{PhiPhi}
\end{eqnarray}
Other Poisson brackets vanish. It is worth notice that
\begin{eqnarray}
\alpha^{ab} &=& \alpha^{ba}, \mathrm{(symmetric)}  \nonumber\\
\rho^{ab} &=& - \rho^{ba}, \mathrm{(antisymmetric)} \nonumber\\
\epsilon^{ab} &=& - \epsilon^{ba}, \mathrm{(antisymmetric)}
\label{syasym}
\end{eqnarray}
Thus the matrix $\Delta_{ij}^{ab}$ between the constraints has the form
\begin{equation}
\Delta_{ij}^{ab} \equiv \{{\Phi_i}^a, {\Phi_j}^b\}_P = 
\begin{bmatrix}
0 & 0 & 0 & - {\alpha}^{ab} \\
0 & 0 & {\alpha}^{ab} & -{\beta}^{ab} \\
0 & -{\alpha}^{ab} & {\rho}^{ab} & -{\gamma}^{ab} \\
{\alpha}^{ab} & {\beta}^{ba} & {\gamma}^{ba} & {\epsilon}^{ab}
\end{bmatrix}
\label{Dijab}
\end{equation}
It is worth note here that $a, b = 1,....,(N-1)$. Hence each element of the matrix $\Delta_{ij}^{ab}$ is a $(N-1)\times (N-1)$ matrix. Thus, the matrix $\Delta_{ij}^{ab}$ is a $4(N-1)\times 4(N-1)$ matrix.
\section{BFFT Formalism}\label{BFFT Formalism}
In this section we will discuss the main results of BFFT technique, which is used to Abelianize the second class constraint systems. 
                             The basic idea behind the scheme is to introduce additional phase space variables $\Theta_m^n$, besides the existing physical degrees of freedom $(q, p)$ of the system such that all the constraints in the extended space of the system are first class. This means that the original constraints and Hamiltonian have to be modified accordingly by putting BFFT-extension terms in them. To achieve this, we will use the results discussed in \cite{IABIVT,RBBN,SG}. Let us consider a set of constraints $(\Phi_n^m, \Lambda_j)$ and an Hamiltonian operator $H$. We know from the Dirac's constraint analysis that second-class constraints of a constrained system satisfy an open algebra. These constraints and Hamiltonian satisfy following algebra
\begin{eqnarray}
\{\Phi_n^m (q, p), \Phi_r^s (q, p)\} \approx \Delta_{nr}^{ms}(q,p) \neq 0,\nonumber\\
\{\Phi_n^m (q, p), \Lambda_r^j (q, p)\} \approx 0,\nonumber\\
\{\Lambda^j (q, p), \Lambda_t (q, p)\} \approx 0,\nonumber\\
\{\Lambda^j (q, p), H (q, p)\} \approx 0
\label{secal}
\end{eqnarray}
$``\approx"$ means that the equality holds on the constraint surface.
The additional fields satisfy the symplectic algebra,
\begin{eqnarray}
\{\Theta_m^n, \Theta_s^r\} = \omega_{ms}^{nr}
\label{pofex}
\end{eqnarray}
where $\omega_{ms}^{nr}$ is a constant quantity and $\det\omega_{ms}^{nr} \neq 0$. The constraints are now defined in terms of auxiliary field $\Theta_m^n$ as 
\begin{eqnarray}
{\tilde\Phi}_n^m  = {\tilde \Phi}_n^m (q, p; \Theta_m^n ),
\label{tithnm}
\end{eqnarray}
This modified constraint satisfies the boundary condition
\begin{eqnarray}
{\tilde \Phi}_n^m(q, p; 0)  =  \Phi_n^m (q, p),
\label{bdccn}
\end{eqnarray}
These modified constraints should satisfy first class constraints algebra. So the Poisson bracket between the constraints are defined as
\begin{eqnarray}
\{{\tilde \Phi}_n^m, {\tilde \Phi}_r^s\} = 0
\label{mfcal}
\end{eqnarray}
The solution of eqn (\ref{mfcal}) can be achieved by considering an expansion of $\Phi_n^m$, as
\begin{eqnarray}
{\tilde \Phi}_n^m = \sum_{k = 0}^{\infty} {\tilde\Phi}_n^{m(k)},
\label{mdccn}
\end{eqnarray}
where ${\tilde\Phi}^{m(k)} \approx O(\Theta^k)$. 
The first order correction in the field is \cite{IABIVT,RBBN,SG}
\begin{eqnarray}
{\tilde\Phi}_n^{m(1)} = X_{nr}^{ms}(q,p)\Theta_{s}^{r}
\label{exton}
\end{eqnarray}
Putting the expression of eqn (\ref{exton}) in eqn (\ref{mfcal}) and using the boundary condition given in eqns (\ref{bdccn}), (\ref{secal}) as well as eqn (\ref{pofex}), we get
\begin{eqnarray}
\Delta_{nr}^{ms} + X_{nc}^{md}\omega_{df}^{ce}X_{re}^{sf} = 0
\label{DeXoX}
\end{eqnarray}
We notice that the eqn (\ref{DeXoX}) does not give a single solution for $X_{ij}^{ab}$, because there is still unknown matrix $\omega_{ab}^{ij}$. We can make choices for $\omega_{ab}^{ij}$ in such a way that the newly defined variables are unconstrained in nature. Using the value of the matrix $\omega_{ab}^{ij}$, we can calculate the possible of $X_{ij}^{ab}$ from the eqn (\ref{DeXoX}). Using the value of $X_{ij}^{ab}$ we can obtain ${\tilde\Theta}_n^{m(1)}$. If ${\Theta_n^m} + {\tilde\Theta}_n^{m(1)}$ is strongly involutive in nature, then series will end otherwise it will continue in the same way till we don't get strongly involutive constraints.
The explicit expression of higher order corrections in the field $\Phi$ is
\begin{eqnarray}
{\tilde\Phi}_n^{m(k+1)} = - \frac{1}{k+2}\Theta_b^c X_{ce}^{bd}\omega_{df}^{eg}B_{gn}^{fm(k)} \quad ; \quad k\geq 1
\label{gexth}
\end{eqnarray}
where $B_{mn}^{ba}$ is defined as
\begin{eqnarray}
B_{rs}^{ba(k)} &=& \sum_{l = 0}^{k}\{{\tilde\Phi}_r^{b(k-l)},{\tilde\Phi}_s^{a(l)}\}_{(q,p)} + \sum_{l = 0}^{k-2}\{{\tilde\Phi}_r^{b(k-l)},{\tilde\Phi}_s^{a(l+2)}\}_{(\Theta)}, \quad k\geq 2\nonumber\\
B_{rs}^{ba(1)} &=& \{{\tilde\Phi}_r^{b(0)},{\tilde\Phi}_s^{a(1)}\}_{(q,p)} - \{{\tilde\Phi}_r^{a(0)},{\tilde\Phi}_s^{b(1)}\}_{(q,p)}
\label{Bmnba}
\end{eqnarray}
In the above expressions, we have defined 
\begin{eqnarray}
X_{mn}^{ab}X_{bc}^{nr} = \omega_{mn}^{ab}\omega_{bc}^{nr} = \delta_m^r\delta_c^a
\label{XmnabX}
\end{eqnarray}
Another important part of the BFFT formalism is that any dynamical variable $f(q, p)$ has also to be modified in the same way as discussed above in order to be strongly involutive with the modified constraints ${\tilde\Phi}_n^m$. Denoting the modified quantity by $f(q,p;\Theta)$, we then have
\begin{eqnarray}
\{{\tilde\Phi}_n^m, {\tilde f}\} = 0
\label{tiThf}
\end{eqnarray}
Apart from that, modified variable ${\tilde f}$ must also satisfy the boundary condition given below,
\begin{eqnarray}
{\tilde f}(q, p; 0)  =  f (q, p)
\label{tifqp}
\end{eqnarray}
To obtain $\tilde f$ as an analogous expansion to eqn (\ref{mdccn}), we will consider
\begin{eqnarray}
{\tilde f} = \sum_{k = 0}^{\infty} {f}^{(k)}      
\label{tifTh}
\end{eqnarray}
where ${\tilde f}^{(k)}$ is also a term which is of the order n in $\Theta^{'}s$. 
The expression in eqn (\ref{tiThf}) above gives us ${\tilde f}^{(1)}$
\begin{eqnarray}
{\tilde f}^{(1)} = - \Theta_n^a\omega_{ab}^{no}X_{om}^{bc}(q,p)\{{\tilde\Phi}_c^m , f \},
\label{fchmt}
\end{eqnarray}
where $\omega_{ab}^{mn}$ and $X_{mn}^{ab}$ are the inverses of $\omega_{mn}^{ab}$ and $X_{ab}^{mn}$.
 
The corrections in the physical variable $f$ can be written in the more general form as,
\begin{eqnarray}
{\tilde f}^{(k+1)} = - \frac{1}{k+1}\Theta_n^a\omega_{ab}^{no}X_{om}^{bc}(q,p){G(f)_c^m}^{(k)},
\label{nocHm}
\end{eqnarray}
where
\begin{eqnarray}
G_a^{b(k)} = \sum_{l = 0}^{k}\{{\tilde\Phi}_r^{b(k-l)}, f^{(l)}\}_{(q,p)} + \sum_{l = 0}^{(k-2)}\{{\tilde\Phi}_r^{b(k-l)}, f^{(l+2)}\}_{(\Theta)} + \{{\tilde\Phi}_r^{b(k+1)}, f^{(1)}\}_{(\Theta)}
\end{eqnarray}
In the similar way, we can find the involutive form of other variables using the BFFT method described above. 

Let us take the initial fields as $q$ and $p$. Then the involutive form of these fields ($\tilde q$ and $\tilde p$), will satisfy the following relations.
\begin{eqnarray}
\{{\tilde \Phi}, {\tilde q}\} = \{{\tilde \Phi}, {\tilde p}\} = 0
\label{minfl}
\end{eqnarray}
Similarly, any function of the physical variables $\tilde q$ and $\tilde p$ will also satisfy the strong involution relation, since
\begin{eqnarray}
\{{\tilde \Theta}, {\tilde F}({\tilde q}, {\tilde p})\} = \{{\tilde \Theta}, {\tilde q}\}\frac{\partial {\tilde F}}{\partial {\tilde q}} + \{{\tilde \Theta}, {\tilde p}\}\frac{\partial {\tilde F}}{\partial {\tilde p}} = 0
\label{tiThF}
\end{eqnarray}
So, if we are taking any dynamical variable in the original phase space, it can be written in involutive form as

\begin{eqnarray}
F (q,p) \rightarrow F({\tilde q}, {\tilde p}) = \tilde F ({\tilde q}, {\tilde p})
\label{Ftqtp}
\end{eqnarray}
It is very much obvious that the initial boundary condition in the BFFT formalism, namely, the reduction of the involutive physical variables to the original physical variables, when the new fields are set to zero, remains preserved.

\section{Construction of the first class constraint Theory}\label{Construction of the first class constraint Theory}
We can easily observe that all the constraints of the theory in eqn (\ref{Phia}) are of second class in nature. To change them in the first class constraints we will introduce $4L$ set of possible BFFT fields $\Theta^{a(1)}, \Theta^{a(2)}, \Theta^{a(3)}, \Theta^{a(4)}$. Here each set of newly introduced fields will correspond to a set of constraints. We will define some relation between these BFFT fields which will help us in the Abelianization of the constraints of the theory. Using the relation between newly introduced fields we will define $\omega$ which will give us the possible solution of the eqn (\ref{DeXoX}). Here we will discuss the Abelianiaztion of constraints and Hamiltonian for the Particle motion on the surface $V_L (1\leq L<N)$ in the Riemann manifold $R_N$ (based on \cite{VPRT}. 

Our choice of Poisson Bracket between the fields $\Theta^{a(1)}, \Theta^{a(2)}, \Theta^{a(3)}, \Theta^{a(4)},\ (a = 1,...,N-1)$ are
\begin{eqnarray}
\{ \Theta^{a(1)}, \Theta^{b(3)} \} &=& I^{ab}, \quad \{ \Theta^{a(2)}, \Theta^{b(4)} \} = I^{ab} 
\label{ph12341}
\end{eqnarray}
where $I^{ab}$ is an $(N-1)\times(N-1)$ unitary matrix.

From the relation between the fields $\Theta$, we can find matrix $\omega_{ab}^{ij}$ as,
\begin{equation}
\omega^{aibj} = 
\begin{bmatrix}
0 & 0 & I^{ab} & 0 \\
0 & 0 & 0 & I^{ab} \\
-I^{ab} & 0 & 0 & 0 \\
0 & -I^{ab} & 0 & 0 
\end{bmatrix}
\label{omabij1}
\end{equation}

Using the matrix $\omega^{aibj}$ defined above and the matrix $\Delta_{ij}^{ab}$ between the constraints in the eqn (\ref{Dijab}), we can calculate the possible value of matrix $X_{ij}^{ab}$. 
Now, using the matrix $X_{ij}^{ab}$, we can write the modified constraints as,
\begin{eqnarray}
{\tilde\Phi}_1^a &=& \Pi^a - \Theta^{a(3)}, \quad {\tilde\Phi}_2^a =  Q^a + \Theta^{a(2)},\quad {\tilde\Phi}_3^a = (P^A - \partial^A {\bar Q}_b\Theta^{b(4)})\partial_A {\bar Q}^a\nonumber\\ 
{\tilde\Phi}_4^a &=& (P^A - \partial^A {\bar Q}_b\Theta^{b(4)}) (P^B - \partial^B {\bar Q}_c\Theta^{c(4)})\partial_A\partial_B{\bar Q}^a\Theta^{b1} - \partial_A {\bar V} \cdot \partial^A {\bar Q}^a \nonumber\\ &+& \lambda_d\partial_A {\bar Q}^a \partial^A {\bar Q}^d  + \partial_A {\bar Q}^a \partial^A {\bar Q}_e \Theta^{e(1)}
\label{mdcnt1}
\end{eqnarray}
It is worth mention here that, all the barred quantities defined in eqn (\ref{mdcnt1}) are function of coordinates $x^k$ and fields $\Theta^{a(2)}$ and will take the form of the original unbarred quantities in the limit $\Theta^{a(2)} \rightarrow 0$. Here, any field $\bar f(x^k, \Theta^{a(2)})$ will be written as \cite{VPRT}
\begin{eqnarray}
{\bar f (x^k, \Theta^{a(2)})} = \sum_{n = 0}^{\infty} \frac{f^{(n)}_a}{n!}\Theta^{a(2)}_{(n)}
\label{baxth}
\end{eqnarray}
also, the partial differentiation of field $\bar f$ wrt. any field $x^k$ can be written as \cite{VPRT}
\begin{eqnarray}
{\bar f_{,i}} = Q_{ai}\{\bar f, \Theta^{a(4)}\}
\label{baxth}
\end{eqnarray}
Then the Poisson bracket between these modified set of constraints is,
\begin{eqnarray}
\{{\tilde\Phi}_i^a, {\tilde\Phi}_j^a\} = 0
\label{tiPia1}
\end{eqnarray}
where $i, j = 1,2,3,4,$ and $a, b = 1, 2,..., N-1$.
We can conclude from eqn (\ref{tiPia1}) that the modified constraints are involutive in nature. Hence we have successfully converted the second class constraints of the theory into first class constraints.

Now, we will construct the involutive Hamiltonian for this system. 

Corrections in the Hamiltonian due to different fields $\Theta$ can be calculated as follows.
We will start it by calculating the inverse of the matrices $\omega^{aibj}$ and $X_{ij}^{ab}$. The inverse of the matrix $\omega^{aibj}$ can be easily written as,
\begin{equation}
\omega_{ij}^{ab} = 
\begin{bmatrix}
0 & 0 & - I^{ab} & 0 \\
0 & 0 & 0 &  - I^{ab}\\
I^{ab} & 0 & 0 & 0 \\
0 & I^{ab} & 0 & 0
\end{bmatrix}
\label{finvom}
\end{equation} 

The total Hamiltonian with corrections due to BFFT field can be written as,
\begin{eqnarray}
{\tilde H} = \frac {1}{2}\cdot (P_A - \partial_A {\bar Q}_b\Theta^{b(4)}) (P^A - \partial^A {\bar Q}_c\Theta^{c(4)}) + \bar V(x) - (\lambda_a + \Theta_{a(1)}) (Q^a(x) + \Theta^{a(2)})
\label{fmodhm}
\end{eqnarray}
Here $\alpha, \beta, \gamma, \rho, \epsilon$ are $(N-1)\times(N-1)$ matrices and $\Theta^{a(i)}(i = 1,2,3,4)$ takes $4(N-1)$ possible values.

Now, by calculating the Poisson bracket between the modified constraints and the Hamiltonian $\tilde H$ it can be easily verified that modified Hamiltonian is involutive in nature.
\begin{eqnarray}
\{\tilde H, \tilde \Phi_i^a\} = 0
\label{tiHtiP}
\end{eqnarray}
where $i = 1,2,3,4$ and $a = 1,2,..., N-1$.

\section{Hamiltonian BRST Quantization}\label{Hamiltonian BRST Quantization}
\subsection{Charge and Symmetry}\label{Charge and Symmetry}
In this section we will construct BRST symmetry for the Particle motion on the surface $V_L (1\leq L<N)$ in the Riemann manifold $R_N$. To construct the we will use the Hamiltonian BRST formalism also called BFV-BRST formalism \cite{MHCT,MH,ESFGV,IABGV,IABESF}. 

In the BFV-BRST technique associated to a general class of system with first class constraints, we introduce two canonical set of ghost and anti-ghost fields  $(C,\bar P)$ with ghost number 1 and -1 respectively and $(P,\bar C)$ with ghost number -1 and 1 respectively with Lagrange multiplier fields $(N,B)$ for each set of constraints.
As there are $4(N-1)$ set of constraints, we will introduce two $4(N-1)$ sets of canonical ghost and anti-ghost fields  $(C^{ka},{\bar P}_k^a)$, $(P^{ka},{\bar C}_k^b)$ and Lagrange multiplier fields $(N^{ka},B_k^a)$.  
These fields and corresponding momenta satisfy following super algebra,
\begin{eqnarray}
\{C^{ka},{\bar P}_l^b\} = \{P^{ka},{\bar C}_l^b\} = \{N^{ka},B_l^b\} =\delta^k_l I^{ab}
\label{ghagha}
\end{eqnarray}
Now, using the BFV formalism we can write the general expression for nilpotent BRST charge, gauge-fixing fermion and BRST invariant Hamiltonian for the system under study as,
\begin{eqnarray}
Q_{BRST} = \int dx^N (C^k_a{\tilde\Omega}_k^a + P^k_a B_k^a)
\label{gnbrch}
\end{eqnarray} 
\begin{eqnarray}
\Psi = \int dx^N({\bar P}_k^a N^k_a + {\bar C}^k_a \chi_k^a )
\label{genpsi} 
\end{eqnarray}
\begin{eqnarray}
H_U = H_P + H_{BF} - \{Q_{BRST}, \Psi\}
\end{eqnarray}
In the BFV-BRST formulation the generating functional is doesn't depend on gauge fixing fermion \cite{ESFGV,IABESF,IABGV,LDFAD,PSENJ}, hence one has freedom to choose it in the convenient form. It is also worth notice here, that gauge-fixing fermion $\chi_k^a$ is Hermitian in nature and of the similar Grassmann parity as of $\tilde\Phi^a$. They also satisfy
\begin{eqnarray}
\det|\{\chi_k^a,\tilde\Phi^l_b\}| \neq 0
\end{eqnarray}

We will apply the general results obtained above to the particle on surface $V^L$ embedded in $R^N$.  
\begin{eqnarray}
S_{eff} = \int dt \big[ P_A\dot x^{A} + \Pi_{\Theta a}^k \dot\Theta^a_k + B_k^a{\dot N}^k_a + {\dot P}^k_a {\bar C}_k^a  + {\dot C}_a^k {\bar P}_k^a - H_P - H_{BF} + \left[Q_{BRST}, \Psi \right] \big] 
\label{eapstk}
\end{eqnarray}
For the system under study, the modified constraints ${\tilde \Phi}_k^a$ have been calculated in eqn (\ref{mdcnt1}).

Our choice of gauge condition $\chi_k^a$ for the motion on the surface $V_L$ in the Riemann manifold $R_N$ is $\Phi_k^a$ in eqn (\ref{Phia}).

Here $\tilde H = H_P + H_{BF}$ is taken as the BFFT modified Hamiltonian for this system obtained in eqn (\ref{fmodhm}).

We will define the canonical brackets for all dynamical variables as
\begin{eqnarray}
 [x^{A}, P_{B}] = {\delta^A}_B;\quad [\Theta^a_i, \Pi^{jb}_{\Theta} ] = [\lambda^a_i, \Pi^{bj} ] = \delta^j_i I^{ab};\quad \{{\bar C}_k^a, {\dot C}^{bl}\} = i\delta^l_k I^{ab}; \quad\{C_k^a, {\dot{\bar C}}^{bl}\} = - i \delta^l_k I^{ab}
\label{cbfadb}
\end{eqnarray} 
Nilpotent BRST transformations corresponding the action in eqn (\ref{eapstk}) can be easily constructed using the relation $S_{\mathrm{BRST}}\Gamma = - [Q_{\mathrm{BRST}}, \Gamma]_{\pm}$. It's relation with infinitesimal BRST transformation $S_{\mathrm{BRST}}$ can be established as $\delta_{\mathrm{BRST}} \ = S_{\mathrm{BRST}} \Gamma \delta \Lambda$. $\delta \Lambda$ defined here is an infinitesimal BRST parameter. Here `$-$' and `$+$' sign are defined for bosonic and fermionic variables respectively. The BRST transformation for the particle motion on general class of Riemann surfaces embedded in higher dimensional Euclidean spaces is written as,
\begin{eqnarray} 
&&S_{\mathrm{BRST}} N^{ak} = P^{ak}, \quad S_{\mathrm{BRST}} {\bar P}^{ak} = {\tilde \Phi}^{ak}, \quad S_{\mathrm{BRST}} {\bar C}^{ak} = -B^{ak}\nonumber\\
&& S_{\mathrm{BRST}} P^A = S_{\mathrm{BRST}}  C^{ak} = S_{\mathrm{BRST}} \Pi^{ak} = S_{\mathrm{BRST}} {P}^{ak} = 0
\label{brtrf1}
\end{eqnarray}
It can be easily verified that these transformations are nilpotent in nature.

Using the expressions for BRST charge $Q_{BRST}$ and gauge-fixing fermion $\Psi$, effective action in eqn (\ref{eapstk}) can be written as
\begin{eqnarray}
S_{eff} = \int dt \big[ P_A{\dot x}^A + \Pi_{\Theta a}^k \dot\Theta_k^a + B_k^a{\dot N}^k_a + {\dot P}^k_a {\bar C}_k^a  + {\dot C}^k_a {\bar P}_k^a - \tilde H - P^k_a{\bar P}_k^a + N_k^a \tilde\Phi^k_a + B_k^a\chi^k_a + {\bar C_k^a}C^k_a \big ]
\label{eleps}
\end{eqnarray}
The generating functional for this effective action can be written  as 
\begin{eqnarray}
Z_\Psi &=& \int [D \varphi] e^{(iS_{eff})}
\label{gnf}  
\end{eqnarray}
The Liouville measure $D\varphi$ for the generating functional is defined as,
\begin{eqnarray}
D\varphi =\prod_i d \Xi_i
\label{Lmes}  
\end{eqnarray}
Here $\Xi_i$ are all dynamical variables $(P_A, x^A, \Pi_{\Theta a}^k, \Theta^a_k, N^k_a, B_k^a, {\bar C}_a^k, P^a_k, C_a^a, {\bar P}^a_k)$ of the theory.
Now performing the integration over ghost and antighost momenta $P^a_k$ and ${\bar P}^a_k$, we will get 
\begin{eqnarray}
&&{Z_\psi} =  \int D \varphi' \exp \big[i\int dt \big[P_A {\dot x}^A  + \Pi_{\Theta a}^k \dot\Theta^a_k + B_k^a\dot N^k_a + \dot C_k^a{\dot {\bar C}^k_a} - \tilde H + N_k^a\tilde \Phi^k_a - C^k_a \bar C_k^a - B_k^a\chi^k_a \big] \big]
\label{gfaioppb}    
\end{eqnarray}
Here $D\varphi'$ is the new path integral measure for the effective action after the performance of integral over the fields $P$ and $\bar P$.
Further performance of integral over fields $B^k_a$ will give us the new effective generating functional as
\begin{eqnarray}
&&{Z_\psi} = \int D \phi'' \exp \big[i\int dt \big[ P_A {\dot x}^A  + \Pi_{\Theta a}^k \dot\Theta^a_k + \dot C_k^a{\dot {\bar C}^k_a} - \tilde H + N_k^a\tilde\Phi^k_a - C^k_a \bar C_k^a - \frac{\{\dot{N_k^a} - \chi_k^a\}^2}{2}  \big] \big]
\label{gfaiopl}    
\end{eqnarray}
where $D\varphi''$ is the new path integral measure which corresponds to all the dynamical variables left out after integration. Now, the new BRST transformation for the modified action in eqn (\ref{gfaiopl}) is written as 
\begin{eqnarray} 
&&S_{\mathrm{BRST}} N^{ak} = {\dot C}^{ak},  \quad S_{\mathrm{BRST}} {\bar C}^{ak} = -{\dot N}^{ak} - \chi^{ak}\nonumber\\
&& S_{\mathrm{BRST}} P^A = S_{\mathrm{BRST}}  C^{ak} = S_{\mathrm{BRST}} B^{ak} = S_{\mathrm{BRST}} {P}^{ak} = 0
\label{brtrf2}
\end{eqnarray}
It is well known in the literature that BRST charges are nilpotent in nature. We also know that the operation of these charges on the states of total Hilbert space gives us the physical subspace of the system. 
\begin{eqnarray}
{Q_{\mathrm{BRST}}} |\mathrm{phys}\rangle = 0, \quad |\mathrm{phys}\rangle \neq {Q_{\mathrm{BRST}}} |....\rangle
\label{bcos}
\end{eqnarray}
which can be further written in more explicit form for this system as,
\begin{eqnarray}
iC_k^a\tilde\Phi^k_a|\mathrm{phys}\rangle  = 0, \quad i\dot{\bar C}_k^a N^k_a |\mathrm{phys}\rangle  = 0
\label{bcost}
\end{eqnarray}
The result of eqn (\ref{bcost}) implies that the first class constraints of the system under study will annihilate the physical subspace of the total Hilbert space of the system.

\section{Consistency with Previous result}\label{Consistency with Previous result}
We can check the consistency of the results of this system in the limit of $N-L = 1$ \cite{VKP}.
In this limit, elements of the matrix will transform as, 
\begin{equation}
\alpha^{ab} \rightarrow \alpha,
\quad \beta^{ab} \rightarrow \beta,
\quad \gamma^{ab} \rightarrow \gamma
\label{albgab}
\end{equation}
From the antisymmetry (\ref{syasym}) 
\begin{equation}
\rho^{ab}\rightarrow 0, \quad \epsilon^{ab}\rightarrow 0
\label{rhepab}
\end{equation}
The inverse matrix elements will transform under this limit as
\begin{equation}
A \rightarrow - \frac{\gamma}{\alpha^2}, \quad B \rightarrow \frac{\beta}{\alpha^2},\quad C \rightarrow \frac{\gamma}{\alpha^2}, \quad D \rightarrow -\frac{\beta}{\alpha^2}, \quad {E, F} \rightarrow 0 
\label{ABCD}
\end{equation}
Here $E, F$ vanishes due to asymmetry.

Also under the transformation $Q^a \rightarrow f(x)$ the modified constraints in eqn (\ref{mdcnt1}) takes the form of modified constraints of $L = N-1$ case,
\begin{eqnarray}
\tilde\Phi_1 &=& \Pi - \Theta^{(3)}\nonumber\\
\tilde\Phi_2 &=& f(x) + \Theta^{(2)} \nonumber\\
\tilde\Phi_3 &=& (\bar P^k - {\partial^k {\bar f(x)}}\Theta^{(4)})\bar{\partial_k f(x)}  \nonumber\\
\tilde\Phi_4 &=& (\bar P^k - \partial^k \bar f(x)\Theta^{(4)})(\bar P^l - \partial^l \bar f(x)\Theta^{(4)})\partial_k \partial_l \bar f(x) - \partial_k \bar V\cdot \partial^k \bar f(x) \nonumber\\ &&+ \lambda \partial_k \bar f(x)\partial^k \bar f(x) + \partial_k \bar f(x)\partial^k \bar f(x)\Theta^1
\label{modcn1}
\end{eqnarray}
The form of modified Hamiltonian in eqn (\ref{fmodhm}) under these transformations is
\begin{eqnarray}
&&\tilde H = \frac {1}{2}\cdot (\bar P_k - \partial_k\bar f(x)\Theta^{(4)}) (\bar P^k - \partial^k\bar f(x)\Theta^{(4)}) + \bar V(x) - (\lambda + \Theta^{(1)})(f(x) + \Theta^{(2)})
\label{modhm1}
\end{eqnarray}
Similary, BRST charge and BRST symmetry can be written in this limit as
\begin{eqnarray}
&&Q_{\mathrm{BRST}} = C^k{\tilde\Phi}_k + P^k B_k\nonumber\\
&&S_{\mathrm{BRST}} N^k = {\dot C}^k,  \quad S_{\mathrm{BRST}} {\bar C}^k = -{\dot N}^k - \chi^k\nonumber\\
&&S_{\mathrm{BRST}} P^a = S_{\mathrm{BRST}} C^k = S_{\mathrm{BRST}} B^k = S_{\mathrm{BRST}} {P}^k = 0
\label{chsym}
\end{eqnarray}
These are the same constraints, Hamiltonian, BRST charge and symmetry which we have obtained in our previous work \cite{VKP}. This shows that all the results obtained here (\ref{mdcnt1},\ref{fmodhm},\ref{brtrf1},\ref{brtrf2}) using BFFT formalism are consistent with the previous results \cite{VKP} in the limit $L \rightarrow (N-1)$.

\section{ Examples of $L (1\leq L < N)$ Dimensional Embedding in $R^N$}\label{Examples of L Dimensional Embedding in $R^N$}
As an example of $L$ dimensional embedding in $R^N$ we will discuss particle on torus knot \cite{VS,PDSG,VKPBPM,VPBM,ASSG}. We will discuss all the important results developed for general system in this case. 
\subsection{Particle on Torus Knot}
Particle on torus knot is a one dimensional surface embedded in three dimensional space. It is a special kind of knot that lies on the surface of un-knotted torus in $R^3$. It is specified by a set of co-prime integers $p$ and $q$. A torus knot of type $(p, q)$ winds $p$ times around the rotational symmetry axis of the torus and $q$ times around a circle in the interior of the torus. The toroidal co-ordinate system is a suitable choice to study this system. Toroidal co-ordinates are related to Cartesian co-ordinates ($x_1, x_2, x_3$) in following ways
\begin{eqnarray}
x_1 = \frac{a \sinh\eta \cos\phi}{\cosh\eta - \cos\theta},\quad x_2 = \frac{a \sinh\eta \sin\phi}{\cosh\eta - \cos\theta}, \quad x_3 = \frac{a \sin\theta}{\cosh\eta - \cos\theta}
\label{ccoo}
\end{eqnarray}
where, $0\leq\eta\leq\infty$, $-\pi\leq\theta\leq\pi$ and $0\leq\phi\leq 2\pi$. A toroidal surface is represnted by some specific value of $\eta$ (say $\eta_0$). Parameters $a$ and $\eta_0$ are written as $a^2 = R^2 - d^2$ and $\cosh\eta_0 = \frac{R}{D}$ where $R$ and $D$ are major and minor radius of torus respectively.

Lagrangian for a particle constrained to move on the surface of torus knot is
\begin{eqnarray}
L = \frac{1}{2} m a^2\frac{{\dot \eta}^2+ {\dot \theta}^2+ {\sinh^2\eta}{\dot\phi}^2}{(\cosh\eta - \cos\theta)^2} + \lambda (p\theta + q\phi)
\label{lpstk}
\end{eqnarray}
where $(r,\theta,\phi)$ are toroidal co-ordinates for toric geometry and $\lambda$ is the Lagrange multiplier.
The canonical Hamiltonian corresponding to the Lagrangian in eqn (\ref{lpstk})is then written as,  
\begin{eqnarray}
H = \frac {(\cosh\eta - \cos\theta)^2}{2m a^2}\left[{p^2_\eta} + {p^2 _\theta}+ \frac{{p^2_\phi}}{{\sinh^2\eta}}\right] - \lambda (p\theta + q\phi)
\label{hpstk}
\end{eqnarray}
where $p_\eta$, $p_\theta, p_\phi$ and ${p}_\lambda$ are the canonical momenta conjugate to the coordinate $\eta$, $\theta$, $\phi$ and $\lambda$ respectively, defined as
\begin{eqnarray}
p_\eta = \frac{ma^2{\dot\eta}}{(\cosh\eta - \cos\theta)^2}, \quad 
p_\theta = \frac{ma^2\dot\theta}{(\cosh\eta - \cos\theta)^2}, \quad
p_\phi = \frac{ma^2\sinh^2\eta\dot\phi}{(\cosh\eta - \cos\theta)^2},\quad
{p}_\lambda \approx 0
\end{eqnarray}
The ${p}_\lambda$ is the primary constraint of the theory.

After inclusion of primary constraint our new Hamiltonian has the form
\begin{equation}
H_T = \frac {(\cosh\eta - \cos\theta)^2}{2m a^2}\left[{p^2_\eta} + {p^2 _\theta}+ \frac{{p^2_\phi}}{{\sinh^2\eta}}\right] - \lambda (p\theta + q\phi) + u p_\lambda
\label{hmt}
\end{equation}
Now, using Dirac's method of Hamiltonian analysis, we will calculate all the possible constraints of the theory as, 
\begin{eqnarray}
&&{\dot p}_\lambda = \{ {p}_\lambda, H_T \}_P = (p\theta + q\phi) \approx 0 \\
&&{\ddot p}_\lambda =  \{ (p\theta + q\phi), H_T \}_P = \frac {(\cosh\eta - \cos\theta)^2}{ma^2} \left[p p_\theta + \frac{q p_\phi}{\sinh^2\eta} \right] \approx 0 \\
&&{p}_\lambda^{(3)} = \{ \frac {(\cosh\eta - \cos\theta)^2}{ma^2} \left[p p_\theta + \frac{q p_\phi}{\sinh^2\eta} \right], H_T \}_P = \frac {(\cosh\eta - \cos\theta)^2}{ma^2}[\{2p p_\theta \sinh{\eta} + \frac{2q p_\phi}{\sinh{\eta}} \nonumber\\ &&- \frac{2q p_\phi\cosh{\eta}(\cosh\eta - \cos\theta)}{\sinh^3{\eta}}\} p_\eta\frac{(\cosh\eta - \cos\theta)}{ma^2} + 2 \sin{\theta}(p p_\theta + \frac{q p_\phi}{\sinh^2\eta})p_\theta\frac{(\cosh\eta - \cos\theta)}{ma^2} \nonumber\\ &&- p\frac{(\cosh\eta - \cos\theta)}{ma^2}\{\sin{\theta}({p^2_\eta} + {p^2 _\theta}+ \frac{p^2_\phi}{\sinh^2\eta}) - \lambda p\} + \lambda \frac{q^2}{{\sinh^2\eta}} ] \approx 0 
\label{ddotp}
\end{eqnarray}
${p}_\lambda^{(4)}$ will vanish and the value of $u$ will be determined from it.
All the constraints can be written as,
\begin{eqnarray}
&&\Phi_1^a = p_{\lambda} \nonumber\\ 
&&\Phi_2^a = Q^a = (p\theta + q\phi) \nonumber\\
&&\Phi_3^a = D Q^a =  \frac {(\cosh\eta - \cos\theta)^2}{ma^2} \left[p p_\theta + \frac{q p_\phi}{\sinh^2\eta} \right] \nonumber\\
&&\Phi_4^a = P^AP^B\partial_A\partial_B Q^a  - \nabla Q^a\cdot{\nabla}(V - \lambda_d Q^d) = D^2 Q^a - \nabla Q^a \cdot{\nabla}\Phi\nonumber\\ &&= \frac{(\cosh\eta - \cos\theta)^2}{ma^2}[\{2p p_\theta \sinh{\eta} + \frac{2q p_\phi}{\sinh\eta} - \frac{2q p_\phi\cosh{\eta}(\cosh\eta - \cos\theta)}{\sinh^3{\eta}}\} p_\eta\frac{(\cosh\eta - \cos\theta)}{ma^2} \nonumber\\ &&+ 2 \sin{\theta}(p p_\theta + \frac{q p_\phi}{\sinh^2\eta})p_\theta\frac{(\cosh\eta - \cos\theta)}{ma^2} - p\frac{(\cosh\eta - \cos\theta)}{ma^2}\{\sin{\theta}({p^2_\eta} + {p^2 _\theta}+ \frac{p^2_\phi}{\sinh^2\eta}) - \lambda p\} \nonumber\\ &&+ \lambda \frac{q^2}{\sinh^2\eta} ]
\label{cnopt}
\end{eqnarray}
Now, the Poisson brackets between the constraints have following values,
\begin{eqnarray}
&&\{\Phi_1^a, \Phi_4^b \}_P = -\nabla Q^a\cdot \nabla Q^b  = - \frac {(\cosh\eta - \cos\theta)^2}{ma^2}[p^2 + \frac{q^2}{\sinh^2\eta}] \equiv - \alpha^{ab}
\label{Phi14}
\end{eqnarray}
\begin{eqnarray}
&&\{\Phi_2^a, \Phi_3^b \}_P = \nabla Q^a\cdot \nabla Q^b = \frac {(\cosh\eta - \cos\theta)^2}{ma^2}[p^2 + \frac{q^2}{\sinh^2\eta}] \equiv  \alpha^{ab}
\label{Phi23}
\end{eqnarray}
\begin{eqnarray}
&&\{\Phi_2^a, \Phi_4^b \}_P = 2\nabla Q^a\cdot (\nabla D Q^b) = 2 \frac {(\cosh\eta - \cos\theta)^3}{m^2a^4} [p\{ p p_\eta \sinh\eta +  \sin{\theta}(p p_\theta + \frac{q p_\phi}{\sinh^2\eta})\}\nonumber\\ && + q\{\frac{q p_\eta}{\sinh{\eta}}(1 - \frac{\cosh{\eta}(\cosh\eta - \cos\theta)}{\sinh^2{\eta}}) + \frac{\sin{\theta}}{\sinh^2\eta}(q p_\theta - p p_\phi)\}] \equiv - \beta^{ab}
\label{Phi24}
\end{eqnarray}
Similar the Poisson bracket between other constraints $(\Phi_3^a, \Phi_4^b)$ (\ref{phi34}), $(\Phi_3^a, \Phi_3^b)$ (\ref{phi33})and $(\Phi_4^a, \Phi_4^b)$ (\ref{phi44}) has been explicitly calculated in the appendix. All these brackets will be nonzero and will be equal to $\gamma^{ab}, \rho^{ab}$ and $\epsilon^{ab}$. Thus the matrix between the constraints will take exactly the same form of matrix $\Delta_{ij}^{ab}$ in eqn ({\ref{Dijab}}).  

As all the constraints of the theory (\ref{cnopt}) are second class, we will follow the method of section IV and introduce four possible fields $\Theta^{a(1)}, \Theta^{a(2)}, \Theta^{a(3)}, \Theta^{a(4)}$ corresponding to each constraint. Relation between these fields will provide us possible value of $\omega^{abij}$.
Our choice of Poisson bracket between the fields will be same as  one taken for the general case.
Hence the matrix $\omega^{abij}$ will have the form of eqn (\ref{omabij1}).

Using the matrix $\omega^{abij}$ and the matrix $\Delta_{ab}^{ij}$, in the eqn (\ref{exton}), one can find many possible value of matrix $X_{ab}^{ij}$. 

Now, applying the results developed in section IV we can calculate the modified constraints as
\begin{eqnarray}
\tilde\Phi_1^a &=& p_\lambda - \Theta^{a(3)}\nonumber\\
\tilde\Phi_2^a &=& (p\theta + q\phi) + \Theta^{a(2)} \nonumber\\
\tilde\Phi_3^a &=& \frac {\big(\cosh\eta - \cos{(\theta - \frac{\Theta^{a(2)}}{2p} )}\big)^2}{ma^2}\big[(p_\theta - p\Theta^{(4a)})p + \frac{(p_\phi - q\Theta^{(4a)})q}{\sinh^2\eta}\big] \nonumber\\
\tilde\Phi_4^a &=& \frac{(\cosh\eta - \cos{(\theta - \frac{\Theta^{a(2)}}{2p})})^2}{ma^2}[\big\{2p (p_\theta - p\Theta^{(4a)}) \sinh{\eta} + \frac{2q (p_\phi - q\Theta^{(4a)})}{\sinh\eta} \nonumber\\ &-& \frac{2q (p_\phi - q\Theta^{(4a)})\cosh{\eta}\big(\cosh\eta - \cos(\theta - \frac{\Theta^{a(2)}}{2p})\big)}{\sinh^3{\eta}}\big\}  p_\eta\frac{(\cosh\eta - \cos(\theta - \frac{\Theta^{a(2)}}{2p}))}{ma^2} \nonumber\\ &+& 2 \sin{(\theta - \frac{\Theta^{a(2)}}{2p})}(p (p_\theta - p\Theta^{(4a)}) + \frac{q (p_\phi - q\Theta^{(4a)})}{\sinh^2\eta})(p_\theta - p\Theta^{(4a)})\frac{(\cosh\eta - \cos(\theta - \frac{\Theta^{a(2)}}{2p}))}{ma^2} \nonumber\\ &-& p\frac{(\cosh\eta - \cos(\theta - \frac{\Theta^{a(2)}}{2p}))}{ma^2}\{\sin{(\theta - \frac{\Theta^{a(2)}}{2p})}({p^2_\eta} + {(p_\theta - p\Theta^{(4a)})^2}+ \frac{(p_\phi - q\Theta^{(4a)})^2}{\sinh^2\eta}) - \lambda p\} \nonumber\\ &+& \lambda \frac{q^2}{\sinh^2\eta} + \big(\frac{(\cosh\eta - \cos(\theta - \frac{\Theta^{a(2)}}{2p}))}{ma^2}p^2+\frac{q^2}{\sinh^2\eta}\big)\Theta^{a(1)}]  
\end{eqnarray}
The Poisson bracket between these modified constraints vanishes which shows that modified constraints are involutive. Hence we have converted the second class constraints of the theory into first class.

Now, we will construct first class Hamiltonian for this system using the results in section V.

The total involutive Hamiltonian for this system will take the form as \cite{VPRT},
\begin{eqnarray}
\tilde H &=& \frac {(\cosh\eta - \cos(\theta - \frac{\Theta^{a(2)}}{2p}))^2}{2m a^2}\left[{p^2_\eta} + {(p_\theta - p\Theta^{(4a)})^2}+ \frac{(p_\phi - q\Theta^{(4a)})^2}{\sinh^2\eta}\right] \nonumber\\ &-& (\lambda + \Theta^{a(1)})(p\theta + q\phi + \Theta^{a(2)}) 
\end{eqnarray}
It can be easily verified that the Hamiltonian $\tilde H$ is involutive by computing it's Poisson bracket with modified constraints of the theory.
\begin{eqnarray}
\{\tilde H, \tilde \Phi_i^a\} = 0
\label{tiHPhtk}
\end{eqnarray}
where $i = 1,2,3,4$.

BRST charge for this first class system can be written using above expression, as
\begin{eqnarray}
Q_{BRST} = i C^i_a\tilde\Phi_i^a + i P^i_a B_i^a
\label{brchtr}
\end{eqnarray}

and corresponding BRST symmetry transformation can be written as
\begin{eqnarray} 
&&S_{\mathrm{BRST}} N^{ak} = P^{ak}, \quad S_{\mathrm{BRST}} {\bar P}^{ak} = {\tilde \Phi}^{ak}, \quad S_{\mathrm{BRST}} {\bar C}^{ak} = -B^{ak}\nonumber\\
&& S_{\mathrm{BRST}} P^A = S_{\mathrm{BRST}}  C^{ak} = S_{\mathrm{BRST}} \Pi^{ak} = S_{\mathrm{BRST}} {P}^{ak} = 0
\label{brsttk}
\end{eqnarray}
This shows that result obtained in section IV is true for any $L(1\leq L < N)$ dimensional surface embedded in $R^N$.

\section{Batalin - Vilkovisky Quantization}\label{Batalin - Vilkovisky Quantization}
In the current section of the manuscript, we are going to discuss the quantization of the Particle motion on the surface $V_L (1\leq L<N)$ in the Riemann manifold $R_N$ using the field-antifield formalism \cite{IABGAV1,IABGAV2,JGJPSS} developed for BFFT systems in \cite{RART}. We will start by introducing $4(N-1)$ set of antifields $\varpi^{k\star}_\mu = (x^{\star}_A, \Theta^{k\star}_a, \lambda^{k\star}_a, C^{k\star}_a)$ corresponding to the fields $\varpi^\mu_k = (x^A, \Theta^a_k, \lambda^a_k, C^a_k)$. Here, fields $ x^A, \Theta^{ak}$ and $\lambda^{ak}$ are bosonic in nature and have ghost number zero whereas the ghost fields $C^{ak}$ are fermionic in nature and have ghost number one. Antifields corresponding to these fields have opposite Grassmann parity and their ghost numbers are given by minus the ghost number of the corresponding fields minus one.

BV-action for this system in terms of fields and antifields is written as
\begin{eqnarray}
S & = & S_0 + \int dt \big[ x^{\star}_A \{ x^A, {\tilde\Phi}_a^k \} C^a_k + {\Theta}^{\star}_{bk} \{ \Theta^{bk}, {\tilde\Phi}_a^l \} C^a_l + \lambda^{k\star}_a {\dot C}^a_k \big ]
\label{acbff}
\end{eqnarray}
where action $S_0$ is defined as
\begin{eqnarray}
S_0 & = &\int dt \big[ P_A{\dot x}^A + \Pi_a^k {\dot\Theta}^a_k - \lambda^a_k {\tilde\Phi}_a^k - {\tilde H}\big ]
\label{ginac}
\end{eqnarray}
Here ${\tilde\Phi^a_i}$ and ${\tilde H}$ are modified constraints and modified Hamiltonian in eqn (\ref{mdcnt1}) and eqn (\ref{fmodhm}) respectively.
The BV-action defined in eqn (\ref{acbff}) satisfies the classical master equation 
\begin{eqnarray}
\frac{1}{2}(S,S) = 0
\label{clmeq}
\end{eqnarray}
where the antibracket between any two dynamical variables $X[\varpi, \varpi^{\star}]$ and $Y[\varpi, \varpi^{\star}]$ is defined as 
\begin{eqnarray}
(X,Y) = \frac{\delta_r X}{\delta \varpi^\mu}\frac{\delta_l Y}{\delta \varpi^{\star}_\mu} - \frac{\delta_r X}{\delta\varpi^{\star}_\mu}\frac{\delta_l Y}{\delta \varpi^\mu }
\label{antib}
\end{eqnarray}
Here, de Witt's notation of sum and integration over intermediary variables will be assumed, whenever necessary.
The BRST differential in is the BV formalism can be introduced using the relation $sX = (X,S)$ for any local functional $X[\varpi, \varpi^{\star}]$ of fields. 
Nilpotentcy of the BRST operator $s$ can be proved using classical master equation and Jacobi identity. So, the BV action satisfying the master equation is equivalent to it's BRST invariance.

To fix gauges, we will extend the Hilbert space to introduce $4(N-1)$ pairs of ghost-antighost fields and corresponding momenta $({\bar C}_a^k, P_a^k)$, $(\bar {C^{k\star}_a}, P^{k\star}_a)$, as well as gauge-fixing fermions $\Psi$. These antifields can be easily eliminated by choosing $\varpi^{\star}_\mu = \frac{\partial \Psi}{\partial \varpi^\mu}$. One of the possible forms of $\Psi$, we can choose for the given system is,
\begin{eqnarray}
\Psi = {\bar C}_a^k \Theta^a_k
\label{gffer}
\end{eqnarray} 
We have liberty to make other possible choices also. 
Now, we will extend the BV action defined above to a nonminimal action,
\begin{eqnarray}
S \rightarrow S_{nm} = S + \int dt P_a^k {\bar C}^{\star a}_k
\label{nnmac} 
\end{eqnarray}
in order to implement the many set of gauge fixing conditions introduced by $\Psi$. Now, the generating functional for gauge-fixed action is defined as
\begin{eqnarray}
{Z_\psi} & = & \int [d\varpi^\mu ][d\omega]^{-\frac{1}{2}} [df]^{-\frac{1}{2}} \exp {\frac{i}{\hbar}S_{nm}[\varpi^\mu, \varpi^{\star}_\mu = \frac{\partial \Psi}{\partial {\varpi^\mu}}]}
\label{vacfn}    
\end{eqnarray}

Now, we will replace the original classical field–antifield action $S$ by some quantum action $\Sigma$ which is expressed as a local functional of fields and antifields and also satisfy a new equation called quantum master equation defined as
\begin{eqnarray}
\frac{1}{2}(\Sigma,\Sigma) - i{\hbar}\Delta \Sigma = 0
\label{qnmeq}
\end{eqnarray}
then the gauge symmetries of the extended action are not obstructed at quantum level. Here $\Delta$ act as an operator and is defined as
\begin{eqnarray}
\Delta \equiv (\frac{\delta_r}{\delta \varpi^\mu})(\frac{\delta_l}{\delta \varpi^{\star}_\mu})
\label{drdpA}
\end{eqnarray}
It is also assumed here that the quantum action $\Sigma$ can be expanded in powers of $\hbar$ in the following manner,
\begin{eqnarray}
\Sigma[\varpi^\mu,\varpi^{\star}_\mu] = S[\varpi^\mu,\varpi^{\star}_\mu] + \sum_{p = 1}^{\infty}{\hbar}^p M_p[\varpi^\mu,\varpi^{\star}_\mu]
\label{qnact}
\end{eqnarray}
The first two term of the quantum master eqn (\ref{qnmeq}) reads as,
\begin{eqnarray}
(S,S) = 0\nonumber\\
(M_1,S) = i\Delta S
\label{qnaeq}
\end{eqnarray}
It can be easily observed that if $\Delta S$ is non-zero and gives a nontrivial result, then there exists some $M_1$ which can be expressed in terms of local fields such that eqn (\ref{qnaeq}) is satisfied.
Using the cohomological arguments, it can be easily shown that the quantum master equation for the first order systems with pure second class constraints converted to first class by the use of the BFFT procedure, can always be solved. BRST transformations for the fields and antifields for the BFFT Abelianized system can be written as follows,
\begin{eqnarray} 
&&S_{\mathrm{BRST}} N^{k\nu} = {\dot C}^{k\nu}, \quad S_{\mathrm{BRST}}  C^{k\nu} = 0, \quad S_{\mathrm{BRST}} {\bar C}^{k\nu} = P^{k\nu}, \quad S_{\mathrm{BRST}} P^{k\nu}  = 0\nonumber\\
&&S_{\mathrm{BRST}} x^{\star}_A = -\frac{\partial S}{\partial x^A}, \quad S_{\mathrm{BRST}} \Theta^{k\star}_\nu = -\frac{\partial^k S}{\partial {\Theta}^\nu},\quad S_{\mathrm{BRST}} N^{k\star}_\nu = {\tilde\Phi^k}_\nu, \quad S_{\mathrm{BRST}} {\bar P}^{{k\star}\nu} =  {\bar C}^{{k\star}\nu}\nonumber\\&& 
S_{\mathrm{BRST}} C^{k\star}_\nu = - x^{\star}_A \{ x^A, {\tilde\Phi}_\nu^k \} - {\Theta}^{i\star}_l \{ \Theta^l_i, {\tilde\Phi}_\nu^k \} - {\dot N}^{k \star}, 
\quad S_{\mathrm{BRST}} \bar{C^{k\star}_\nu} = 0 
\label{bvbrst}
\end{eqnarray}
The symmetry transformations obtained in the eqn (\ref{bvbrst}) are identical to the one obtained in eqn (\ref{brtrf1}) using BFV formalism. Also, we can easily show on the basis of argument given in \cite{RART} that the enlarged symmetries due to the compensating fields (BFFT variables) are non-anomalous in nature. These BFFT fields also plays very significant role at the quantum level because of the existence of a counterterm, by modifying the expectation values of the relevant physical quantities.

\section{Result and Discussion}\label{Result and Discussion}
The BRST symmetry for a particle moving in a curved space $V_{L}(1\leq L < N) $ embedded in a Euclidean space $R_N$ is investigated in both Hamiltonian and Lagrangian formalism. All the constraints of the system have been calculated using Dirac's Hamiltonian analysis. Using the algebra of constraints, we have found that all the constraints of the system are second class. To construct a gauge invariant theory, we have used the BFFT technique. Using this technique, all the second class constraints of the system are converted into first class constraints and corresponding Hamiltonian is also constructed explicitly. Using the involution of Hamiltonian with first class constraints, this Hamiltonian is shown to be first class. In the limit of $\Theta \rightarrow 0$ the constraints and Hamiltonian returns to original second class constraints and Hamiltonian. Now, using this gauge invariant system, we have constructed BRST charge, symmetries and the BRST invariant action. For constructing BRST symmetry from the first class constraint system, BFV formalism is used. These BRST charges acting on the stetes of the total Hilbert space, will annihilate the physical subspace of it. From it, we have deduced that first class constraints operating on total Hilbert space of the system will annihilate its physical subspace which can be used as a physicality criteria for the BRST invariant system. We have shown that the general results derived here for any surface embedded in $R^N$ is consistent with the results of previous work \cite{VKP}. We have also discussed particle motion on the torus knot surface as an example of this kind of system. In this example, we have explicitly calculated all the constraints of this system and converted them to first class constraints. It has been found that diagonal elements $\rho$ and $\epsilon$ of matrix (of the Poisson's bracket between the constraints) are nonzero if we take two different torus knot systems. In the limit of $a\rightarrow b$, the $\rho, \epsilon$ will vanish and we will achieve the commutative case of \cite{VKP}. We have also constructed the first class Hamiltonian, BRST charge and symmetry for this system. It has been shown that all the results deduced for the general system are consistent with this system. At the end we have discussed Batalin - Vilkovisky quantization of this system based on BFFT formalism. Here also we have explicitly calculated the BRST symmetry for the general system which is consistent with the symmetry derived from Hamiltonian formalism. This again proves the equivalence between the Hamiltonian and Lagrangian formalism. Recently a more general technique of Lagrangian Abelianization has been developed {\cite{SLL,VAASLL}}. It will be interesting to apply this technique to the motion in Riemann manifold.

\section{Appendix}\label{Appendix}
The calculation of Poisson bracket between some of the constraints for particle on the torus knot model is straightforward but cumbersome process. We have calculated these brackets explicitly here.

The Poisson bracket between constraints $(\Phi_3^a, \Phi_4^b)$ is written as,
\begin{eqnarray}
&&\{\Phi_3^a, \Phi_4^b \}_P = 2\nabla (D Q^a)\cdot \nabla (D Q^b) - \nabla Q^a\cdot \nabla \Phi_4^b = 2\frac {(\cosh\eta - \cos\theta)^4}{m^3a^6}\{ \sinh\eta(p p_\theta + \frac{q\phi}{\sinh^2\eta}) \nonumber\\ && - \frac{q p_\phi\cosh\eta (\cosh\eta - \cos\theta)}{\sinh^3{\eta}}\}\cdot \{p p_\theta \sinh\eta + \frac{q p_\phi}{\sinh\eta} - \frac{q p_\phi\cosh\eta (\cosh\eta - \cos\theta)}{\sinh^3{\eta}} - p p_\eta \sin\theta\} \nonumber\\ &&+ 4 \frac{(\cosh\eta - \cos\theta)^4}{m^3a^6} \sin\theta (p p_\theta + \frac{q\phi}{\sinh^2\eta})\{p p_\eta \sinh\eta + \sin\theta (p p_\theta + \frac{q\phi}{\sinh^2\eta})\}- p\frac{(\cosh\eta - \cos\theta)^2}{ma^2}\nonumber\\ &&\big[2\sin\theta\frac{(\cosh\eta - \cos\theta)}{ma^2}\cdot\{ \{2p p_\theta \sinh{\eta} + \frac{2q p_\phi}{\sinh{\eta}} - \frac{2q p_\phi\cosh{\eta}(\cosh\eta - \cos\theta)}{\sinh^3{\eta}}\} p_\eta\frac{(\cosh\eta - \cos\theta)}{ma^2} \nonumber\\ &&+ 2 \sin{\theta}(p p_\theta + \frac{q\phi}{\sinh^2\eta})p_\theta\frac{(\cosh\eta - \cos\theta)}{ma^2} - p\frac{(\cosh\eta - \cos\theta)}{ma^2}\{\sin{\theta}({p^2_\eta} + {p^2 _\theta}+ \frac{p^2_\phi}{\sinh^2\eta}) - \lambda p\} \nonumber\\ &&+ \lambda \frac{q^2}{{\sinh^2\eta}} \} +\frac{(\cosh\eta - \cos\theta)^2}{ma^2} \{ -2q \sin\theta p_\phi\frac{\cosh\eta}{\sinh^3\eta}\cdot p_\eta\frac{(\cosh\eta - \cos\theta)}{ma^2}\} + \{2p p_\theta \sinh{\eta}  \nonumber\\ &&+ \frac{2q p_\phi}{\sinh{\eta}} - \frac{2q p_\phi\cosh{\eta}(\cosh\eta - \cos\theta)}{\sinh^3{\eta}}\}\frac{p_\eta \sin\theta}{ma^2} + 2p_\theta \cos\theta (p p_\theta + \frac{q p_\phi}{\sinh^2\eta})\frac{(\cosh\eta - \cos\theta)}{ma^2}\nonumber\\ &&+ 2\sin\theta (p p_\theta + \frac{q p_\phi}{\sinh^2\eta})\frac{p_\theta\sin\theta}{ma^2} - \frac{p_\theta\sin\theta}{ma^2}\{\sin\theta({p^2_\eta} + {p^2 _\theta}+ \frac{p^2_\phi}{\sinh^2\eta}) - \lambda p\} - p\cos\theta\nonumber\\ && \frac{(\cosh\eta - \cos\theta)}{ma^2}({p^2_\eta} + {p^2 _\theta}+ \frac{p^2_\phi}{\sinh^2\eta})\big]\equiv - \gamma^{ab}
\label{phi34}
\end{eqnarray}

Poisson bracket $(\Phi_3^a, \Phi_3^b)$ is written as,
\begin{eqnarray}
&&\{\Phi_3^a, \Phi_3^b \}_P = \nabla (D Q^a)\cdot \nabla Q^b - \nabla Q^a\cdot \nabla (D Q^b)\nonumber\\ && = \{2\frac{(\cosh\eta^a - \cos\theta^a)}{ma^2}(p {p_{\theta^a}} + \frac{q\phi^a}{\sinh^2\eta^a})\cdot p'\frac{(\cosh\eta^b - \cos\theta^b)^2}{ma^2} \nonumber\\ &&- p\frac{(\cosh\eta^a - \cos\theta^a)^2}{ma^2}\cdot2\frac{(\cosh\eta^b - \cos\theta^b)}{ma^2}(p' p_{\theta^b} + \frac{q'\phi^b}{\sinh^2\eta^b})\}\equiv \rho^{ab}
\label{phi33}
\end{eqnarray}
Similarly for $(\Phi_4^a, \Phi_4^b)$ can be written as,
\begin{eqnarray}
&&\{\Phi_4^a, \Phi_4^b \}_P = 2\left[\nabla \Phi_4^a\cdot \nabla (D Q^a) - \nabla \Phi_4^b\cdot \nabla (D Q^a)\right] = 2\frac{(\cosh\eta^a - \cos\theta^a)}{ma^2}\sinh\eta^a[\{2p p_{\theta^a} \sinh{\eta^a} \nonumber\\&& + \frac{2q p_{\phi^a}}{\sinh\eta^a} - \frac{2q p_{\phi^a}\cosh{\eta^a}(\cosh\eta^a - \cos\theta^a)}{\sinh^3{\eta^a}}\} p_{\eta^a}\frac{(\cosh\eta^a - \cos\theta^a)}{ma^2} + 2 \sin{\theta^a}(p p_{\theta^a} + \frac{q\phi^a}{\sinh^2{\eta^a}})p_\theta\nonumber\\&&\frac{(\cosh\eta^a - \cos\theta^a)}{ma^2} - p\frac{(\cosh\eta^a - \cos\theta^a)}{ma^2}\{\sin{\theta^a}({p^2_{\eta^a}} + {p^2 _{\theta^a}}+ \frac{p^2_{\phi^a}}{\sinh^2{\eta^a}}) - \lambda^a p\} + \lambda^a \frac{q^2}{\sinh^2{\eta^a}} ] \nonumber \\&&+ \frac{(\cosh\eta^a - \cos\theta^a)^2}{ma^2}[\{2p p_{\theta^a} \cosh{\eta^a} - \frac{4q p_{\phi^a} \cosh{\eta^a}}{\sinh^2{\eta^a}} - \frac{2q p_{\phi^a}(\cosh\eta^a - \cos\theta^a)}{\sinh^2{\eta^a}} \nonumber\\&& + 6\frac{q p_{\phi^a}\cosh^2{\eta^a}(\cosh\eta^a - \cos\theta^a)}{\sinh^4{\eta^a}}\} p_{\eta^a} \frac{(\cosh\eta^a - \cos\theta^a)}{ma^2} + \frac{p_{\eta^a} \sinh\eta^a}{ma^2} \{2p p_{\theta^a} \sinh{\eta^a} + \frac{2q p_{\phi^a}}{\sinh{\eta^a}} \nonumber\\&& - \frac{2q p_{\phi^a}\cosh{\eta^a}(\cosh\eta^a - \cos\theta^a)}{\sinh^3{\eta^a}}\} - 4 \sin\theta^a(\frac{2q p_{\phi^a}\cosh\eta^a}{\sinh^3{\eta^a}})\frac{(\cosh\eta^a - \cos\theta^a)}{ma^2}p_{\theta^a} + 2\sin\theta^a(p p_{\theta^a} \nonumber\\&&+ \frac{q p_{\phi^a}}{\sinh^2\eta^a}) \frac{\sinh\eta^a}{ma^2}p_\theta - p\frac{\sinh\eta^a}{ma^2}\{\sin{\theta^a}({p^2_{\eta^a}} + {p^2 _{\theta^a}} + \frac{p^2_{\phi^a}}{\sinh^2{\eta^a}}) - \lambda^a p\} + 2p\frac{(\cosh\eta^a - \cos\theta^a)}{ma^2} \nonumber\\&&\{\sin\theta^a\frac{p^2_{\phi^a} \cosh\eta^a}{\sinh^3{\eta^a}}\} - 2 \lambda^a \frac{q^2 \cosh\eta^a}{\sinh^3{\eta^a}}] \cdot 2\frac{(\cosh\eta^b - \cos\theta^b)^3}{m^2a^4}\{p p_{\theta^b} \sinh\eta^b + \frac{q p_{\phi^b}}{\sinh\eta^b} \nonumber\\&&- \frac{q p_{\phi^b}\cosh\eta^b (\cosh\eta^b - \cos\theta^b)}{\sinh^3{\eta}}  - p p_{\eta^b} \sin\theta^b\} - 2\frac{(\cosh\eta^a - \cos\theta^a)^3}{m^2a^4}\{p p_{\theta^a} \sinh\eta^a + \frac{q p_{\phi^a}}{\sinh\eta^a} \nonumber\\&&- \frac{q p_{\phi^a}\cosh\eta^a (\cosh\eta^a - \cos\theta^a)}{\sinh^3{\eta^a}} - p p_{\eta^a} \sin\theta^a \} \cdot 2\frac{(\cosh\eta^b - \cos\theta^b)}{ma^2}\sinh\eta^b[\{2p p_{\theta^b} \sinh{\eta^b} \nonumber\\&&+ \frac{2q p_{\phi^b}}{\sinh\eta^b} - \frac{2q p_{\phi^b}\cosh{\eta^b}(\cosh\eta^b - \cos\theta^b)}{\sinh^3{\eta^b}}\} p_{\eta^b}\frac{(\cosh\eta^b - \cos\theta^b)}{ma^2} + 2 \sin{\theta^b}(p p_{\theta^b} + \frac{q\phi^b}{\sinh^2{\eta^b}})p_{\theta^b}\nonumber\\&&\frac{(\cosh\eta^b - \cos\theta^b)}{ma^2}- p\frac{(\cosh\eta^b - \cos\theta^b)}{ma^2}\{\sin{\theta^b}({p^2_{\eta^b}} + {p^2 _{\theta^b}} + \frac{p^2_{\phi^b}}{\sinh^2{\eta^b}}) - \lambda^b p\} + \lambda^b \frac{q^2}{\sinh^2{\eta^b}} ] \nonumber\\&&+ \frac{(\cosh\eta^b - \cos\theta^b)^2}{ma^2}[\{2p p_{\theta^b} \cosh{\eta^b} - \frac{4q p_{\phi^b} \cosh{\eta^b}}{\sinh^2{\eta^b}} - \frac{2q p_{\phi^b}(\cosh\eta^b - \cos\theta^b)}{\sinh^2{\eta^b}} \nonumber\\&&+ 6\frac{q p_{\phi^b}\cosh^2{\eta^b}(\cosh\eta^b - \cos\theta^b)}{\sinh^4{\eta^b}}\} p_{\eta^b} \frac{(\cosh\eta^b - \cos\theta^b)}{ma^2} + \frac{p_{\eta^b} \sinh\eta^b}{ma^2}\{2p p_{\theta^b} \sinh{\eta^b} + \frac{2q p_{\phi^b}}{\sinh{\eta^b}} \nonumber\\&&- \frac{2q p_{\phi^b}\cosh{\eta^b}(\cosh\eta^b -\cos\theta^b)}{\sinh^3{\eta^b}}\} - 4 \sin\theta^b(\frac{2q p_{\phi^b}\cosh\eta^b}{\sinh^3{\eta^b}})\frac{(\cosh\eta^b - \cos\theta^b)}{ma^2}p_{\theta^b} + 2\sin\theta^b(p p_{\theta^b} \nonumber\\&&+ \frac{q p_{\phi^b}}{\sinh^2\eta^b}) \frac{\sinh\eta^b}{ma^2}p_{\theta^b} - p\frac{\sinh\eta^b}{ma^2}\{\sin{\theta^b}({p^2_{\eta^b}} + {p^2 _{\theta^b}} + \frac{p^2_{\phi^b}}{\sinh^2{\eta^b}}) - \lambda^b p\} + 2p\frac{(\cosh\eta^b - \cos\theta^b)}{ma^2} \nonumber\\&&\{\sin\theta\frac{p^2_{\phi^b} \cosh\eta^b}{\sinh^3{\eta^b}}\}  - 2 \lambda^b \frac{q^2 \cosh\eta^b}{\sinh^3{\eta^b}}] + [2\sin\theta^a\frac{(\cosh\eta^a - \cos\theta^a)}{ma^2}\cdot\{ \{2p p_{\theta^a} \sinh{\eta^a} + \frac{2q p_{\phi^a}}{\sinh{\eta^a}} \nonumber\\&&- \frac{2q p_{\phi^a}\cosh{\eta^a}(\cosh\eta^a - \cos\theta^a)}{\sinh^3{\eta^a}}\} p_{\eta^a}\frac{(\cosh\eta^a - \cos\theta^a)}{ma^2} + 2 \sin{\theta^a}(p p_{\theta^a} + \frac{q\phi^a}{\sinh^2\eta^a}) p_{\theta^a}\nonumber\\&&\frac{(\cosh\eta^a - \cos\theta^a)}{ma^2} - p\frac{(\cosh\eta^a - \cos\theta^a)}{ma^2}\{\sin{\theta^a}(p^2_{\eta^a} + p^2 _{\theta^a}+ \frac{p^2_{\phi^a}}{\sinh^2\eta^a}) - \lambda^a p\} + \lambda^a \frac{q^2}{{\sinh^2\eta^a}} \} \nonumber\\&&+\frac{(\cosh\eta^a - \cos\theta^a)^2}{ma^2} \{ -2q \sin\theta^a p_{\phi^a}\frac{\cosh\eta^a}{\sinh^3\eta^a}\cdot p_{\eta^a}\frac{(\cosh\eta^a - \cos\theta^a)}{ma^2}\} + \{2p p_{\theta^a} \sinh{\eta^a} + \frac{2q p_{\phi^a}}{\sinh{\eta^a}} \nonumber\\ &&- \frac{2q p_{\phi^a}\cosh{\eta^a}(\cosh\eta^a - \cos\theta^a)}{\sinh^3{\eta^a}}\} \frac{p_{\eta^a} \sin\theta^a}{ma^2} + 2p_{\theta^a} \cos\theta^a (p p_{\theta^a} + \frac{q p_{\phi^a}}{\sinh^2\eta^a})\frac{(\cosh\eta^a - \cos\theta^a)}{ma^2} \nonumber\\&&+ 2\sin\theta^a (p p_{\theta^a} + \frac{q p_{\phi^a}}{\sinh^2\eta^a})\frac{p_{\theta^a}\sin\theta^a}{ma^2} - \frac{p_{\theta^a}\sin\theta^a}{ma^2}\{\sin\theta^a(p^2_{\eta^a} + p^2 _{\theta^a} + \frac{p^2_{\phi^a}}{\sinh^2\eta^a}) - \lambda^a p\} - p\cos\theta^a \nonumber\\&&\frac{(\cosh\eta^a - \cos\theta^a)}{ma^2}(p^2_{\eta^a} + p^2 _{\theta^a} + \frac{p^2_{\phi^a}}{\sinh^2\eta^a})\big]\cdot 2\frac{(\cosh\eta^b - \cos\theta^b)^3}{m^2a^4}\{p p_{\eta^b}\sinh\eta^b + p p_{\theta^b} + \frac{q p_{\phi^b}\sin\theta^b}{\sinh^2\eta^b}\} \nonumber\\&&- 2\frac{(\cosh\eta^a - \cos\theta^a)^3}{m^2a^4}\{p p_{\eta^a}\sinh\eta^a + p p_{\theta^a} + \frac{q p_{\phi^a}\sin\theta^a}{\sinh^2\eta^a}\}\cdot \big[2\sin\theta^b\frac{(\cosh\eta^b - \cos\theta^b)}{ma^2}\cdot\{ \{2p p_{\theta^b} \nonumber\\ &&\sinh{\eta^b} + \frac{2q p_\phi^b}{\sinh\eta^b} - \frac{2q p_{\phi^b}\cosh{\eta^b}(\cosh\eta^b - \cos\theta^b)}{\sinh^3{\eta^b}}\} p_\eta^b\frac{(\cosh\eta^b - \cos\theta^b)}{ma^2} + 2 \sin{\theta^b}(p p_{\theta^b} + \frac{q\phi^b}{\sinh^2\eta^b})\nonumber\\ &&p_\theta^b\frac{(\cosh\eta^b - \cos\theta^b)}{ma^2} - p\frac{(\cosh\eta^b - \cos\theta^b)}{ma^2}\{\sin{\theta^b}(p^2_{\eta^b} + p^2 _{\theta^b}+ \frac{p^2_{\phi^b}}{\sinh^2\eta^b}) - \lambda^b p\} + \lambda^b \frac{q^2}{{\sinh^2\eta^b}} \} \nonumber\\ &&+\frac{(\cosh\eta^b - \cos\theta^b)^2}{ma^2} \{ -2q \sin\theta^b  p_{\phi^b}\frac{\cosh\eta^b}{\sinh^3\eta^b}\cdot p_{\eta^b}\frac{(\cosh\eta^b - \cos\theta^b)}{ma^2}\} + \{2p p_{\theta^b} \sinh{\eta^b}  + \frac{2q p_{\phi^b}}{\sinh{\eta^b}} \nonumber\\ &&- \frac{2q p_{\phi^b}\cosh{\eta^b}(\cosh\eta^b - \cos\theta^b)}{\sinh^3{\eta^b}}\}\frac{p_{\eta^b} \sin\theta^b}{ma^2} + 2p_{\theta^b} \cos\theta^b (p p_{\theta^b} + \frac{q p_{\phi^b}}{\sinh^2\eta^b})\frac{(\cosh\eta^b - \cos\theta^b)}{ma^2} \nonumber\\ && + 2\sin\theta^b (p p_{\theta^b} + \frac{q p_{\phi^b}}{\sinh^2\eta^b})\frac{p_{\theta^b}\sin\theta^b}{ma^2} - \frac{p_{\theta^b}\sin\theta^b}{ma^2}\{\sin\theta^b(p^2_{\eta^b} + p^2 _{\theta^b} + \frac{p^2_{\phi^b}}{\sinh^2\eta^b}) - \lambda^b p\}  \nonumber\\ &&- p\cos\theta^b \frac{(\cosh\eta^b - \cos\theta^b)}{ma^2}(p^2_{\eta^b} + p^2 _{\theta^b}+ \frac{p^2_{\phi^b}}{\sinh^2\eta^b})\big] \equiv \epsilon^{ab}
\label{phi44}
\end{eqnarray}

\end{document}